\documentclass[prd,twocolumn,preprintnumbers,showpacs,amssymb,superscriptaddress]{revtex4}

\usepackage{amsmath}
\usepackage{graphicx}
\usepackage{epsfig}
\usepackage{dcolumn}
\usepackage{bm}
\usepackage{multirow}
\usepackage{rotating}
\usepackage[usenames]{color}

\usepackage{epsfig}
\newcommand{\newc}{\newcommand}

\newc{\nn}{\noindent}
\newc{\non}{\nonumber}

\def\gsim{\lower0.5ex\hbox{$\:\buildrel >\over\sim\:$}}
\def\lsim{\lower0.5ex\hbox{$\:\buildrel <\over\sim\:$}}

\def\ptmiss{\not\!\!{p_T}}

\def\rpv{\not\!\!{R_p}}
\def\L{\not\!\!{L}}
\def\r7{\begin{sideways} $\sqrt{s}=7$ TeV \end{sideways}}

\def\lm2pi{\lambda^{{\prime\prime}^{ijk}}}

\newcommand{\bea}{\begin{eqnarray}}
\newcommand{\eea}{\end{eqnarray}}
\def\beq{\begin{equation}}
\def\eeq{\end{equation}}
\def\wt{\widetilde}

\newcommand{\ntrl}[1]{\widetilde{\chi}^0_#1}

\newcommand{\bsq}[1]{\widetilde{b}_#1}
\newcommand{\tsq}[1]{\widetilde{t}_#1}
\newcommand{\ns}[1]{S^0_#1}
\newcommand{\nps}[1]{P^0_#1}
\newcommand{\cs}[1]{S^\pm_#1}

\newcommand{\chpm}[1]{\widetilde{\chi}^{\pm}_#1}

\begin{document}
\vspace{-0.2cm}
\title{Unusual Higgs boson signal in R-parity violating nonminimal
supersymmetric models at the LHC}
\author{Priyotosh Bandyopadhyay\footnote{Present address: 
Helsinki Institute of Physics and Helsinki University
Gustaf H\"allstr\"omin Katu 2, POB-64, FIN-00014, Helsinki, Finland.}}
\email{priyotosh.bandyopadhyay@helsinki.fi}
\affiliation{Korea Institute for Advanced Study, Hoegiro 87 (207-43 Cheongnyangni-dong),
Seoul 130-722, Korea}
\author{Pradipta Ghosh}%
\email{tppg@iacs.res.in}
\affiliation{Department of Theoretical Physics, 
Indian Association for the Cultivation of Science, 2A and 2B Raja
S.C. Mullick Road, Kolkata 700 032, India
}
\author{Sourov Roy$^2$}%
\email{tpsr@iacs.res.in}
%
\begin{abstract}
We predict an unconventional background free signal of the Higgs boson in $R$-parity 
violating nonminimal supersymmetric models at the Large Hadron Collider (LHC). 
The signal comprises dilepton plus four hadronic jets and two large displaced 
vertices. The displaced leptons and jets are coming from the decay of the lightest 
supersymmetric particle (LSP), which is predominantly a gauge-singlet neutrino. A 
pair of such LSPs can couple to a Higgs boson, created via gluon fusion. We have analyzed 
two cases - one corresponding to the tree-level Higgs boson mass and another with the 
one-loop corrected mass of the Higgs boson. A reliable Higgs mass reconstruction using 
this signal can lead to discovery at the LHC with center-of-mass energy $\sqrt{s}=14~\rm{TeV}$ 
and 5 fb$^{-1}$ of integrated luminosity $(\mathcal{L})$. Even at $\sqrt{s}=7~\rm{TeV}$ 
and $\mathcal{L}$ = 5 fb$^{-1}$, a reasonable number of events are expected. Besides, 
mass reconstruction of a gauge-singlet LSP can provide an estimate of the seesaw scale.
\end{abstract}

\preprint{KIAS-P10048}


\maketitle

\newpage

\section{Introduction}\label{intro}
Supersymmetric theories have commendable success in taming the ill behaved quadratic 
divergences of the mass of the hitherto unseen standard model (SM) Higgs boson.
Recently, inception of the LHC experiment at CERN has given fresh impetus to the 
hope of discovering the Higgs boson and weak scale supersymmetry.

In this paper, we propose a prodigious signal for the Higgs boson in supersymmetry, 
having a pair of leptons and four hadronic jets along with large displaced vertices 
$(\gsim 2 ~m)$. With the aid of these displaced vertices, all of the SM 
backgrounds can be effaced. Displaced vertices arising from minimal supersymmetric 
standard model (MSSM) with $R_p$ violation $(\rpv)$ (see review \cite{rpv-review}) are usually 
much smaller \cite{LSP-DL}, and thus hardly mimic this signal. Furthermore, the imprint of 
this signal is different from that of the cosmic muons which have definite entry and exit 
point in the detector. So this is apparently a clean signal and a discovery, thus, is 
definite even with small number of signal events.

This unusual signal would produce visible tracks in the muon chamber of the CMS or ATLAS 
detector (electrons will be absorbed in the iron yoke of muon-chamber), along with some 
accompanying noise (from the associated hadronic jets). These tracks are isolated
tracks produced by stand-alone muons without any matching tracks in the inner detector.
It is indeed difficult for the conventional triggers to work for this specific signal; 
rather this asks for a dedicated special trigger which we believe is a challenging task 
for experimentalists. Nevertheless, in addition to this final state topology,
one can also have the usual global muon signatures which produce continuous matching tracks 
in the detector, starting from the inner tracker to the muon chamber, but with smaller displaced 
vertices. The associated electrons and hadronic jets also leave their signatures in the inner 
tracker and calorimeters. These events can be easily triggered with conventional triggers.


\section{The Model}\label{model}

The underlying model contains $\rpv$ and, is non-minimal in nature with SM gauge
singlet right-handed neutrino superfields $(\hat \nu^c_i)$. The right sneutrinos 
$(\wt \nu^c_i)$ through their vacuum expectation values provide 
a solution \cite{Mun-Lop-1,Mun-Lop-2} to the $\mu$-problem \cite{Kim-Nilles}. Besides, 
the right-handed neutrinos $(\nu^c_i)$ together with $\rpv$ are instrumental 
for light neutrino mass generation \cite{Pg1,Porod-Bartl,Munoz-Lopez_CP,Pg2},
consistent with three flavour global fit \cite{Valle_nu_dat}.
This model is known as the $\mu\nu$SSM \cite{Mun-Lop-1,Mun-Lop-2}.

The superpotential of $\mu\nu$SSM is given by,
\bea
W &=& W^\prime + \epsilon_{ab} Y^{ij}_\nu \hat H^b_u\hat L^a_i\hat \nu^c_j
-\epsilon_{ab} \lambda^i\hat \nu^c_i\hat H^a_d\hat H^b_u\nonumber \\
&+&\frac{1}{3}\kappa^{ijk}\hat \nu^c_i\hat \nu^c_j\hat \nu^c_k,
\label{superpotential}
\eea
where $W^\prime$ is the MSSM superpotential without the $\mu$-term. 
Lepton number violation $(\L)$ by odd unit(s)
in the last two terms of Eq.(\ref{superpotential}) triggers $\rpv$. 
The fourth term of Eq.(\ref{superpotential}) apart from excluding the
massless axion \cite{Ellis-etal}, generates $\rm{TeV}$ scale right-chiral 
neutrino {\it{Majorana}} masses. Bilinear terms are forbidden in $W$ by 
an imposed $Z_3$ symmetry. The associated problem of domain wall formation 
\cite{Domain-Wall} from spontaneous breaking of this $Z_3$ symmetry
can be ameliorated through known methods \cite{Domain-Wall-soln}.
Scalar and fermion sectors for $\mu\nu$SSM are addressed  
in Refs. \cite{Mun-Lop-2,Pg1,Porod-Bartl,Pg2}.
%



The neutralino LSP, $\ntrl1$ in $\mu\nu$SSM can be predominantly 
$(\gsim 70\%)$ $\nu^c$-like (also known as a {\it{singlino}} LSP). 
$\ntrl1$ being singlet, $\ntrl1 \ntrl1 Z$ or $\ntrl1 q \widetilde q$ 
couplings \cite{Pg2} are vanishingly small, which in turn results in very small 
cross-section for direct $\ntrl1$ pair production. On the contrary, the third 
term of Eq.(\ref{superpotential}) may produce a large $\ntrl1 \ntrl1 S^0_i$ 
\cite{Pg2} coupling with $\lambda \sim$ $\cal{O}$ $(1)$, where $S^0_i$ are the scalar 
states. With the chosen set of parameters (see Table \ref{mass-spectrum}) 
we obtained $S^0_4 \equiv h^0$, where $h^0$ is the lightest Higgs boson of MSSM. 
In addition, with heavy squark/gluino masses as indicated in Table \ref{mass-spectrum}
for different benchmark points, production of a singlino LSP through
cascade decays is suppressed. In the backdrop of such a scenario, production of $h^0$ in 
the gluon fusion channel followed by the decay process $h^0 \to \ntrl1 \ntrl1$ will be the 
leading production channel for the singlino LSP at the LHC. We want to emphasize here that
for the first part of our analysis we choose to work with the tree level mass of the lightest 
CP-even Higgs boson $(S^0_4 \equiv h^0)$ of the $\mu\nu$SSM. With loop corrections the Higgs 
boson mass can be much higher \cite{Mun-Lop-2,Porod-Bartl}. For loop corrected Higgs boson mass, 
the process $h^0\to\ntrl1\ntrl1$ can produce heavy singlinolike LSPs with smaller decay 
lengths \cite{Porod-Bartl}. However, our general conclusions will not change for a singlino 
LSP in the mass range $40-60$ GeV. We will present a similar analysis with the one-loop 
correction in the neutral scalar sector in detail, to justify our statement.


\section{Scenario-I}
\label{S-I}

In this section we present a detailed analysis of the unconventional signal
mentioned in the introduction, with the tree level mass spectra of $\mu\nu$SSM.
A set of four benchmark points (BPs) used for collider studies compatible with 
neutrino data \cite{Valle_nu_dat}, up to one-loop level analysis of neutrino masses, 
and mixing \cite{Pg2} are given in Table \ref{mass-spectrum}. Here we have shown 
the physical mass spectra and the values of only two parameters, namely,
$\mu$ and $\tan\beta$. Other parameters are not shown here. However, note that
these are sample points and similar spectra can be obtained in a reasonably
large region of the parameter space even after satisfying all the constraints 
from neutrino experiments.
%
\begin{table}[ht]
\scriptsize
\caption{\label{mass-spectrum}
$\mu$-parameter, $\rm{tan}\beta$ and relevant mass spectrum (GeV) 
for the chosen benchmark points. 
$m_{\widetilde{\chi}^{\pm}_{1,2,3}} \equiv ~m_{e,\mu,\tau}$.
}
\begin{ruledtabular}
\begin{tabular}{ c  c  c  c  c}
  & BP-1 & BP-2 & BP-3 & BP-4 \\ \hline
$\mu$ & 177.0 & 196.68 & 153.43 & 149.12 \\
$\tan\beta$ & 10 & 10 & 30 & 30 \\
$m_{h^0}~(\equiv m_{\ns4})$ & 91.21 & 91.63 & 92.74 & 92.83    \\ 
$m_{\ns1}$ & 48.58 & 49.33 & 47.27 & 49.84   \\
$m_{\nps2}$ & 47.21 & 49.60 & 59.05 & 49.45   \\ 
$m_{\cs2}$ & 187.11 & 187.10 & 187.21 & 187.21   \\ 
$m_{\bsq1}$ & 831.35 & 831.33 & 830.67 & 830.72  \\ 
$m_{\bsq2}$ & 875.03 & 875.05 & 875.72 & 875.67   \\ 
$m_{\tsq1}$ & 763.41 & 763.63 & 761.99 & 761.98   \\ 
$m_{\tsq2}$ & 961.38 & 961.21 & 962.46 & 962.48   \\ 
$m_{\ntrl1}$ & 43.0 & 44.07 & 44.20 & 44.24   \\ 
$m_{\ntrl2}$ & 55.70 & 57.64 & 61.17 & 60.49   \\ 
$m_{{\chpm4}}$ & 151.55 & 166.61 & 133.69 & 130.77 
\end{tabular}
\end{ruledtabular}
\end{table}
 
For the set of specified benchmark points, we observe, the process $h^0\to \ntrl1 \ntrl1$ to
be one of the dominant decay modes of $h^0$ (branching fraction within $35$-$65\%$), while
the process $h^0 \to b \bar{b}$ remains the main competitor. In Table \ref{mass-spectrum} only the 
third generation squark masses are shown. 

The mass of $\ntrl1$ ($m_{\ntrl1}$) for the chosen points is always less than $m_W$ 
(see Table \ref{mass-spectrum}). For such a light $\ntrl1$, at the tree level only 
three body decay modes are allowed. General three body final states are,
\bea
\ntrl1 &\to& b \bar{b}\nu_k,~\ell^+_i \ell^-_j\nu_k,
~q_i\bar{q_i}\nu_k,~q_i\bar{q}^{\prime}_j\ell^{\mp}_k,~\nu_i\bar{\nu}_j\nu_k,
\label{2-3-body-decays}
\eea
where $i,j,k$ are flavor indices. We choose
the specific decay mode $\ntrl1 \to q_i\bar{q}^{\prime}_j\ell^\pm_k$ 
to yield a signal $pp\to 2\ell + 4j + X$ in the final state. The dilepton have 
same sign on $50\%$ occurrence since $\ntrl1$ is a Majorana particle.
Detection of these leptons and jets can lead to reliable mass 
reconstruction for $\ntrl1$ and Higgs boson in the absence of missing
energy in the final state. There is one more merit of this analysis; 
i.e., invariant mass reconstruction for a singlino LSP can give 
us an estimation of the seesaw scale, since the right-handed neutrinos are 
operational in light neutrino mass generation through a TeV scale seesaw 
mechanism \cite{Mun-Lop-1,Pg1}. For the chosen benchmark points 
$Br(\ntrl1 \to q_i\bar{q}^{\prime}_j\ell^\pm_k)$ lies within $8-10\%$.
However, concerning the real experimental ambience, extra jets can arise from 
initial state radiation and final state radiation. Likewise semileptonic 
decays of quarks can accrue extra leptons. Also from the experimental point of view 
one cannot have zero missing $p_T$ in the final state. With this set of information 
we optimize our chosen signal as
%
\bea
(n_{j} \ge 4) + (n_{\ell} \geq 2) + (\ptmiss \le 30 ~{\rm{GeV}}),
\label{signal-choice}
\eea
%
where $n_{j(\ell)}$ represents the number of {\it{jets(leptons}}).

It should be noted that, similar final states can appear from the decay of 
heavier scalar or pseudoscalar states in the model. Obviously, their production cross 
section will be smaller compared to $h^0$ and the invariant mass distribution 
(some other distributions also) should be different in those cases.

{\tt PYTHIA (version 6.4.22) \cite{PYTHIA}} has been used for the purpose of event generation. 
The corresponding mass spectrum and decay branching fractions 
are fed to {\tt PYTHIA} by using the SLHA interface \cite{Skands-SLHA}. Subsequent decays of the
produced particles, hadronization and the collider analysis were performed using {\tt PYTHIA}. 
We used {\tt CTEQ5L} parton distribution function (PDF) \cite{CTEQ-PDF} for the analysis. The 
renormalization/factorization scale $Q$ was chosen to be the parton level center-of-mass energy, 
$\sqrt{\hat{s}}$. We also kept initial state radiation, final state radiation 
and multiple interaction on for the analysis. The 
production cross-section of $h^0$ via gluon fusion channel for different benchmark points 
(Table \ref{mass-spectrum}) is shown in Table \ref{tabcross}.
\begin{table}[ht]
\scriptsize
\caption{\label{tabcross}
Hard scattering cross-section in fb for the process
$gg \to h^0$ 
for PDF CTEQ5L with 
$Q = \sqrt{\hat{s}}$. 
}
\begin{ruledtabular}
\begin{tabular}{ c  c  c  c  c}
  & BP-1 & BP-2 & BP-3 & BP-4 \\ \hline
$\sqrt{s}=7$ TeV &6837 &7365 &6932 &6948  \\ 
$\sqrt{s}=14$ TeV &23150 &25000 &23580 &23560 
\end{tabular}
\end{ruledtabular}
\end{table}

%
We have used {\tt PYCELL}, the toy calorimeter simulation provided in
{\tt PYTHIA}, with the following criteria:

\noindent
I. The calorimeter coverage is $\rm |\eta| < 4.5$ and the segmentation is
given by $\rm\Delta\eta\times\Delta\phi= 0.09 \times 0.09 $ which resembles
a generic LHC detector.

\noindent
II. $\Delta R \equiv \sqrt{(\Delta\eta)^{2}+(\Delta\phi)^{2}} = 0.5$
        has been used in cone algorithm for jet finding.

\noindent
III. $p_{T,min}^{jet} = 10$ GeV.

\noindent
IV. No jet matches with a hard lepton in the event.

In addition, the following set of standard kinematic cuts
were incorporated throughout:

\noindent
1. $p_T^{\ell} \geq 5$ GeV and $\rm |\eta| _{\ell} \le 2.5$,

\noindent
2. $|\eta| _{j}\leq 2.5$, $\Delta R_{\ell j} \geq 0.4$,
$\Delta R_{\ell\ell}\geq 0.2,~$

\noindent
where $\Delta R_{\ell j}$ and $\Delta R_{\ell \ell}$ measure the lepton-jet and 
lepton-lepton isolation, respectively. Events with isolated leptons, having $p_T\ge 5$ 
GeV are taken for the final state analysis.

\begin{figure}[ht]
\includegraphics[width=7.80cm]{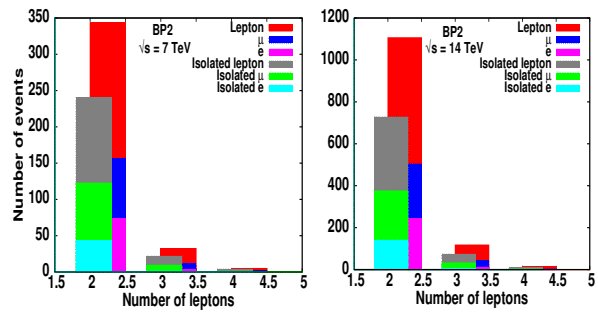}
\caption{Lepton multiplicity  distribution of signal
for $\sqrt{s} = 7$ and $14$ TeV with $1~\rm{fb}^{-1}$
of integrated luminosity. For more details, see text.}
\label{lmlt}
\end{figure}

One of the striking features in $\mu\nu$SSM is that certain ratios of branching fractions of 
the LSP decay modes are correlated with the neutrino mixing angles \cite{Pg1,Porod-Bartl}. 
This implies $n_{\mu} > n_{e}$ in the final state. Fig.\ref{lmlt} shows the lepton 
multiplicity distribution for inclusive $\geq 2\ell$ ($\geq 2 \mu~+\geq 2 e+1\mu,1e$)
and exclusive ($\geq 2 \mu$, $\geq 2 e$) for BP-2, without the signal criteria 
[Eq.(\ref{signal-choice})]. Muon dominance of the higher histograms 
(without any isolation cuts) continues to the lower ones even after the application 
of $\Delta R_{\ell j},~\Delta R_{\ell \ell}$ cuts. Consequently, we observe that the 
correlation between $n_e$ and $n_{\mu}$ also appears in the lower 
histograms (Fig.\ref{lmlt}) with a ratio $\sim$ $1$:$3$.

We present number of events for final state signal (Eq. \ref{signal-choice}) 
in Table \ref{tabevents} both for $\sqrt{s}=7$ and $14$ TeV for $\mathcal{L} = 5~ {\rm fb}^{-1}$, 
without a cut on the actual $\ntrl1$ decay position. 
The average decay length for a singlinolike LSP is determined by
the LSP mass as well as by a set of parameters
$(\lambda,\kappa,v^c,Y^{ii}_\nu,v'_i)$ so
that the constraints on neutrino masses and mixing are satisfied.
Here $v^c$ and $v'_i$ stand for the vacuum expectation values of
the right- and left-handed sneutrino fields. The average decay length 
increases for smaller LSP mass and for a singlinolike LSP with a mass
$\sim$ 20 GeV, the average decay length comes out to be $\sim$ 100 meters \cite{Porod-Bartl}. 
For such a choice, a large fraction of the LSPs will decay outside the detector. We have not 
considered such possibilities in our analysis. The average decay lengths for all the four 
benchmark points are shown in Table \ref{decay-length}. It is interesting to note from 
Table \ref{tabevents} that the correlation between $n_e$ and $n_{\mu}$ in the final
state is still well maintained, similar to what was shown in the lower histograms of 
Fig.\ref{lmlt} even with the final state signal topology [Eq.(\ref{signal-choice})]. 

\begin{table}[ht]
\scriptsize
\caption{\label{tabevents}
Expected number of events of signals for $\mathcal{L} =$ $5 ~\rm{fb}^{-1}$ 
for $\sqrt{s}=7$ and $14~\rm{TeV}$.
}
\begin{ruledtabular}
\begin{tabular}{ c   c  c  c  c c}
$\sqrt{s}$ & signal & BP-1 & BP-2 & BP-3 & BP-4 \\ \hline
& $\ge 4j + \geq 2\ell + \ptmiss \le 30~\rm{GeV}$
&181  &153    &170 &173  \\ 
7 & $\ge 4j + \geq 2\mu + \ptmiss \le 30~\rm{GeV}$ 
&100  &85    &97  &100 \\ 
$\rm{TeV}$ & $\ge 4j + \geq 2e + \ptmiss \le 30~\rm{GeV}$ 
&27  &23   &21  & 23\\ 
& $\ge 4j + 1e +1\mu + \ptmiss \le 30~\rm{GeV}$
&54  &46  &52 &50\\ \hline
 & $\ge 4j + \geq 2\ell + \ptmiss \le 30~\rm{GeV}$
&1043 &878  & 951& 929 \\
14 & $\ge 4j + \geq 2\mu + \ptmiss \le 30~\rm{GeV}$ 
&580 &463  & 533 & 513\\ 
$\rm{TeV}$ & $\ge 4j + \geq 2e + \ptmiss \le 30~\rm{GeV}$
&160 &139  & 121&129  \\ 
& $\ge 4j + + 1e +1\mu + \ptmiss \le 30~\rm{GeV}$
&306 &279 &  300&290 
\end{tabular}
\end{ruledtabular}
\end{table}

\begin{figure}[ht]
\centering
\includegraphics[width=5.40cm,angle=-90]{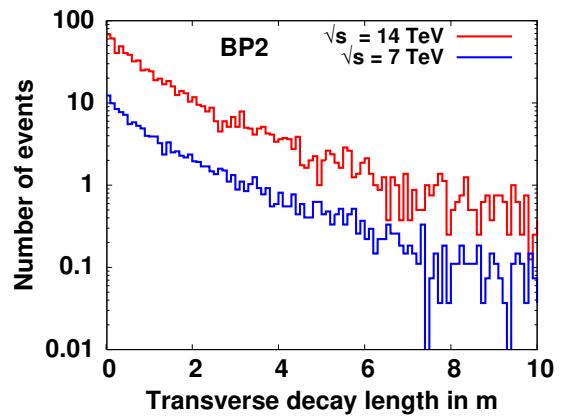}
\caption{Transverse decay length distribution of $\ntrl1$ for $\sqrt{s} = $ 7 
and 14 TeV with BP-2 for a typical detector size $\sim$ $10~m$ with $\mathcal{L}=$
$5 ~\rm{fb}^{-1}$. Minimum bin size is $10~cm$. 
The signal is given by Eq.(\ref{signal-choice}).}
\label{Decay-Length}
\end{figure}

\begin{table}[ht]
\scriptsize
\caption{\label{decay-length}
Average decay lengths $(L)$ for a singlino LSP for different benchmark
points.}
\begin{ruledtabular}
\begin{tabular}{ c  c  c  c  c}
  & BP-1 & BP-2 & BP-3 & BP-4 \\ \hline
$L$ (meter) & 5.20 & 4.23 & 5.20 & 5.35  
\end{tabular}
\end{ruledtabular}
\end{table}
The entries of Table \ref{tabevents} suffer a diminution with
a cut on the $\ntrl1$ decay position. Table \ref{tabevents-cut} represents the number
of signal events in different ranges of the transverse decay length $(L_T)$ for BP-2, 
after applying the cuts. 
Other benchmark points show similar
behaviour. The number of events with decay length (see Table \ref{tabevents-cut}) 
near the muon chamber at CMS is approximately $\sim 10\%$ of 
the total number of events. 
\begin{table}[ht]
\scriptsize
\caption{\label{tabevents-cut}
Number of signal events for $\mathcal{L} =$ $5 ~\rm{fb}^{-1}$ 
for $\sqrt{s}=7$ and $14~\rm{TeV}$ at different ranges of the decay length for BP-2
with $1~cm<L_{T_1}\le50~cm$, $50~cm<L_{T_2}\le3~m$ and $3~m<L_{T_3}\le6~m$.
$L_{T_i}$s are different transverse decay lengths.}
\begin{ruledtabular}
\begin{tabular}{ c c  c  c  c}
 &  &  & No. of events & \\
$\sqrt{s}$ & signal & $L_{T_1}$ & $L_{T_2}$ & $L_{T_3}$\\ \hline
& $\ge 4j + \geq 2\ell + \ptmiss \le 30~\rm{GeV}$
&45  &69   &17 \\ 
7 & $\ge 4j + \geq 2\mu + \ptmiss \le 30~\rm{GeV}$ 
&27  &38  &11   \\ 
$\rm{TeV}$ & $\ge 4j + \geq 2e + \ptmiss \le 30~\rm{GeV}$ 
&6  &10  &2  \\ 
& $\ge 4j + 1e +1\mu + \ptmiss \le 30~\rm{GeV}$
&12  &21 &4 \\ \hline
 & $\ge 4j + \geq 2\ell + \ptmiss \le 30~\rm{GeV}$
&234 &373  &98 \\
14 & $\ge 4j + \geq 2\mu + \ptmiss \le 30~\rm{GeV}$ 
&128 &218  &58 \\ 
$\rm{TeV}$ & $\ge 4j + \geq 2e + \ptmiss \le 30~\rm{GeV}$
&37 &45 &16  \\ 
& $\ge 4j + + 1e +1\mu + \ptmiss \le 30~\rm{GeV}$
&69 &113  &24
\end{tabular}
\end{ruledtabular}
\end{table}
However, being free of backgrounds even this can lead to  
discovery at 14 TeV run of the LHC with $\mathcal{L} =$ 5 fb$^{-1}$. At 7 TeV the 
situation looks much less promising and much higher luminosity is required for 
discovering such an event. There exists reasonable number of events $(\sim 30\%)$ 
with decay length in between {\it{$1$ cm {\rm{and}} $50$ cm}} (see Table \ref{tabevents-cut} 
and Fig.\ref{Decay-Length}). As stated earlier these events can leave their signs 
in inner tracker and calorimeters as well as can produce tracks in the muon chamber. 
Besides, about $\sim 40\%$ of signal events appear in the range of
$50~cm$ to $3~m$ which may or may not leave traces in tracker and calorimeters depending 
on the decay length but will definitely produce visible tracks in the muon chamber.
Thus we can have novel final states observed not only in the muon 
chambers of the LHC detectors but also in the inner tracker and calorimeters.
Therefore combining these different signatures new discoveries are envisaged.
A very small fraction of events $(\sim 4\%)$ do have a 
decay length greater than the size of the CMS detector (For ATLAS
this fraction is somewhat smaller) which will yield the conventional 
missing energy signature. 
%

\begin{figure}[ht]
\centering
\includegraphics[width=7.80cm]{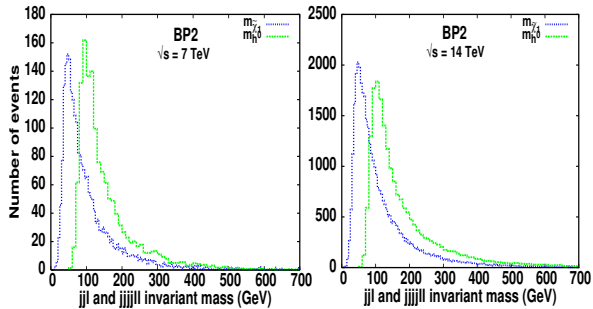}
\caption{Invariant mass distribution for $\ntrl1$ $(jj\ell )$,
as well as the Higgs boson $(jjjj\ell\ell )$. Plots are shown for
$\sqrt{s} = 7$ and $14$ TeV with $1~\rm{fb}^{-1}$
of integrated luminosity.
The Number of events for reconstructing
$m_{\ntrl1}$ for $\sqrt{s} = 7(14)$ TeV
are scaled by a multiplicative factor $4(7)$.}
\label{invariant-masses}
\end{figure}

Results of invariant mass reconstruction
for $\ntrl1$ and $h^0$ for BP-2 are shown in Fig.\ref{invariant-masses}.
We choose $jj\ell$ invariant mass $M(jj\ell)$ for $m_{\ntrl1}$ reconstruction.
Reconstruction of $m_{h^0}$ was achieved through $M(jjjj\ell\ell)$,
invariant mass of $jjjj\ell\ell$ [see Eq.(\ref{signal-choice})].
We take the jets and leptons from the window of 
$35\, \rm{GeV}\, \leq M(jj\ell)\leq \, 45\, \rm{GeV}$ to construct 
$M(jjjj\ell\ell)$. Even a narrow window like this cannot kill all the 
combinatorial backgrounds. As a corollary, effect of combinatorial background 
for $m_{\ntrl1}$ reconstruction ($^4C_2$ for $j$ and $^2C_1$ for $\ell$)
also causes long tail for Higgs mass distribution.
%


\section{Scenario-II}\label{S-II}
In this section we present a similar analysis like scenario-I, but with the
inclusion of one-loop radiative correction in the neutral scalar sector.
Mass of the lightest doubletlike Higgs boson increases as an effect
of adding the one-loop correction in the scalar sector and consequently the 
process $h^0\to\ntrl1\ntrl1$ can yield a heavier singlinolike LSP, compared to 
the benchmark points presented in Table \ref{mass-spectrum}. Also a heavier 
singlino LSP will give rise to a smaller decay length and thus offer a better 
chance of detecting this unusual signal at a collider experiments. All of these 
features will be explored in further details in the subsequent paragraphs. For the 
one-loop corrections to the neutral scalar sector of the $\mu\nu$SSM we consider 
the dominant contributions only, coming from the quarks and squarks in the 
loop\cite{Mun-Lop-2}. The $A_0$ and $B_0$ functions \cite{Veltman-tHooft,Passarino-Veltman} 
appearing in the loop calculations are chosen according to 
Ref. \cite{Pierce:1996zz}. A set of two benchmark points with one-loop corrected 
neutral scalar masses are shown in Table \ref{mass-spectrum-2}.
\begin{table}[ht]
\scriptsize
\caption{\label{mass-spectrum-2}
$\mu$-parameter, $\rm{tan}\beta$ and relevant mass spectrum (GeV) for 
benchmark points with one-loop corrected scalar masses. 
$m_{\widetilde{\chi}^{\pm}_{1,2,3}} \equiv ~m_{e,\mu,\tau}$
and $m_{h^0}~\equiv m_{\ns4}$.
}
\begin{ruledtabular}
\begin{tabular}{ c  c  c}
  & BP-5 & BP-6 \\ \hline
$\mu$ & 137.19 & 145.08 \\
tan$\beta$ & 30 & 10 \\
$m_{h^0}$  & 115.01 & 119.36    \\ 
$m_{\ns1}$ & 61.11 & 61.07   \\
$m_{\ns2}$ & 61.11 & 61.07   \\
$m_{\ns3}$ & 61.21 & 62.25   \\
$m_{\nps2}$ & 69.87 & 66.59   \\ 
$m_{\cs2}$  & 187.22 & 187.12   \\ 
$m_{\bsq1}$ & 830.85 & 831.37  \\ 
$m_{\bsq2}$ & 875.55 & 875.00   \\ 
$m_{\tsq1}$ & 761.93 & 763.06   \\ 
$m_{\tsq2}$ & 962.51 & 961.66   \\ 
$m_{\ntrl1}$ & 52.72 & 54.60   \\ 
$m_{\ntrl2}$ & 73.71 & 74.90   \\ 
$m_{{\chpm4}}$ & 120.85 & 125.53
\end{tabular}
\end{ruledtabular}
\end{table}

Since we consider effect of one-loop corrections only to the neutral scalar sector 
and not for the neutral pseudoscalar sector, for these set of benchmark points neutrino 
data \cite{Valle_nu_dat} are satisfied only for tree level neutrino masses and mixing. 
A complete analysis of one-loop corrected neutrino masses and mixing with scalar as well as
pseudoscalar Higgs boson masses at the one-loop level is beyond the scope of the present paper.
The trilinear couplings $A_u$ and $A_d$ have been chosen accordingly \cite{Mun-Lop-2}, 
in order to obtain different mass eigenvalues of the lightest doubletlike neutral Higgs boson. 
Another important observation with the inclusion of one-loop radiative corrections
to the scalar sector of $\mu\nu$SSM is that for the doubletlike Higgs $(h^0\equiv\ns4)$
the following two body decays $h^0\to S^0_iS^0_j,P^0_aP^0_b$ with $i,j=1,2,3$
and $a,b=2,3,4$ become kinematically more accessible. Thus unlike $h^0\to b \bar{b}$ for 
scenario-I, these processes can now become the dominant competitors to achieve a reasonable 
branching ratio for the process $h^0\to\ntrl1\ntrl1$. 
However, in the present analysis we have avoided such a situation. A pair of benchmark points 
has been chosen such that the two-body decays like $h^0\to S^0_iS^0_j, P^0_aP^0_b$ are kinematically
disfavoured. These are presented in Table \ref{mass-spectrum-2}. For these benchmark points (BP$5$ and BP$6$), 
the process $h^0\to b \bar{b}$ remains the main competitor to achieve a large $Br(h^0\to \ntrl1 \ntrl1)$. 
Three body decays of $h^0$ into $Zf\bar{f}$ or $W^\pm f_1 f_2$ (where $f,f_1,f_2$ are either leptons or quarks) 
share very small branching ratios \cite{Porod-Bartl} when two body decays are kinematically allowed. 
For the two benchmark points shown in Table \ref{mass-spectrum-2} the singlino purity 
is set to be $\gsim 60\%$.

As argued earlier, with one-loop corrected Higgs boson mass, the process $h^0\to \ntrl1\ntrl1$
can yield heavy singlinolike LSPs with smaller decay lengths which are shown in Table \ref{decay-length2}. 
It is clear from Table \ref{decay-length2}, that for $m_{\ntrl1} \sim 50-55$ GeV, the decay lengths are 
$\sim$ $\cal{O}$ $(2-2.5$ m). The singlino LSP mass as shown in Table \ref{mass-spectrum-2} are higher 
compared to those shown in Table \ref{mass-spectrum}. In general, the average decay length can vary from 
$2-8$ m depending on the LSP mass in the range of $40-55$ GeV. It is to be noted that unlike scenario-I, 
here the branching ratio for the process $h^0\to\ntrl1\ntrl1$ lies within $45\%-85\%$, whereas 
$Br(\ntrl1 \to q_i\bar{q}^{\prime}_j\ell^\pm_k)$ lies in the same range as in scenario-I.

As discussed earlier, even lighter singlinolike LSP ($m_{\ntrl1}$ $\sim 20$ GeV) 
can be achieved with smaller values of the trilinear coupling $\kappa$
in the superpotential. However, the associated average decay length will be
very large $\sim$ ${\cal O}$ (100) m \cite{Porod-Bartl} and a large fraction of the LSPs will decay 
outside the detector. In addition to a very large decay length, smaller $\kappa$ values will 
make decays like $h^0 \to S^0_i S^0_j, P^0_a P^0_b$ kinematically possible, with lighter scalar 
and pseudoscalar states. These decays, when allowed are usually the dominant decay 
modes for $h^0$ which lead to the
different final state. Thus, combining the two scenarios (with tree-level and one-loop corrected
Higgs boson mass), we conclude that the discovery of the signal discussed in this paper works in 
the range of LSP mass $\sim$ 40-60 GeV.

\begin{table}[ht]
\scriptsize
\caption{\label{decay-length2}
Average decay lengths $(L)$ for a singlino LSP for different benchmark
points with one-loop corrected scalar masses.}
\begin{ruledtabular}
\begin{tabular}{ c  c  c}
  & BP-5 & BP-6 \\ \hline
$L$ (meter) & 2.65 & 2.10  
\end{tabular}
\end{ruledtabular}
\end{table}

The approach of collider analysis is exactly the same as that of scenario-I. The 
production cross-section of one-loop corrected $h^0$ via gluon fusion channel for 
different benchmark points (Table \ref{mass-spectrum-2}) is given in Table \ref{tabcross-2}.
It is evident from Table \ref{tabcross} and Table \ref{tabcross-2} that the 
hard scattering cross-section for $h^0$ production through gluon fusion at the LHC
suffers diminution with the inclusion of one-loop radiative corrections.
\begin{table}[ht]
\scriptsize
\caption{\label{tabcross-2}
Hard scattering cross-section in fb for the process
$gg \to h^0$ for PDF CTEQ5L with 
$Q = \sqrt{\hat{s}}$ with one-loop corrected scalar masses. 
}
\begin{ruledtabular}
\begin{tabular}{ c  c  c}
  & BP-5 & BP-6 \\ \hline
$\sqrt{s}=7$ TeV &4211.04 &3759.31 \\ 
$\sqrt{s}=14$ TeV &15135.68   &13712.05 
\end{tabular}
\end{ruledtabular}
\end{table}

Also we present the number of signal events with a cut on $\ntrl1$ decay position 
in Table \ref{tabevents-cut-2} for BP-5, which is exactly similar to that of Table \ref{tabevents-cut} 
but with one-loop corrected CP-even scalar masses. 

\begin{table}[ht]
\scriptsize
\caption{\label{tabevents-cut-2}
Number of signal events for $\mathcal{L} =$ $5 ~\rm{fb}^{-1}$ 
for $\sqrt{s}=7$ and $14~\rm{TeV}$ at different ranges of the decay length for BP-5
with $1~cm<L_{T_1}\le50~cm$, $50~cm<L_{T_2}\le3~m$ and $3~m<L_{T_3}\le6~m$.
$L_{T_i}$s are different transverse decay lengths 
with one-loop corrected scalar masses.}
\begin{ruledtabular}
\begin{tabular}{ c c  c  c  c}
 &  &  & No. of events & \\
$\sqrt{s}$ & signal & $L_{T_1}$ & $L_{T_2}$ & $L_{T_3}$\\ \hline
& $\ge 4j + \geq 2\ell + \ptmiss \le 30~\rm{GeV}$
&71  &101   &15 \\ 
7 & $\ge 4j + \geq 2\mu + \ptmiss \le 30~\rm{GeV}$ 
&45  &64  &9   \\ 
$\rm{TeV}$ & $\ge 4j + \geq 2e + \ptmiss \le 30~\rm{GeV}$ 
&8  &11  &0.4  \\ 
& $\ge 4j + 1e +1\mu + \ptmiss \le 30~\rm{GeV}$
&18  &26 &6 \\ \hline
 & $\ge 4j + \geq 2\ell + \ptmiss \le 30~\rm{GeV}$
&371 &523  &86 \\
14 & $\ge 4j + \geq 2\mu + \ptmiss \le 30~\rm{GeV}$ 
&221 &315  &52 \\ 
$\rm{TeV}$ & $\ge 4j + \geq 2e + \ptmiss \le 30~\rm{GeV}$
&50 &72 &12  \\ 
& $\ge 4j + + 1e +1\mu + \ptmiss \le 30~\rm{GeV}$
&99 &138  &22
\end{tabular}
\end{ruledtabular}
\end{table}
Since the average decay length is smaller for a heavier neutralino originating from
the decay of $h^0$ with one-loop corrected mass, now almost $\sim 35\%$ of the
total number of events appears within $1-50~{cm}$ and roughly $\sim 50\%$ of
the total number of signal events appears within $50~cm-3~m$. Integrating the signatures 
of this signal in the two above mentioned range of decay lengths, one can observe either 
tracks in the tracker and calorimeters along with conventional global muon signature (LSP decay 
length in the range $1-50~{cm}$) or tracks in the muon chamber without matching tracks in the 
tracker and calorimeters (for LSP decay length close to $3~m$). In the case of LSP decay lengths between 50 cm and
2 m, one can have signatures in the electromagnetic calorimeter and/or hadronic calorimeter 
(depending on the actual decay position) along with the muon 
tracks in the muon chamber. The fraction of total number of events appearing completely in the muon chamber 
(LSP decay length between $3$ and $6~m$) with stand alone muon signature is slightly less ($\sim 8\%$) compared to scenario-I, 
but still reasonable enough to lead to new discovery particularly with the $\sqrt{s}=14$ TeV run of LHC. 

\begin{figure}[ht]
\centering
\includegraphics[width=5.40cm,angle=-90]{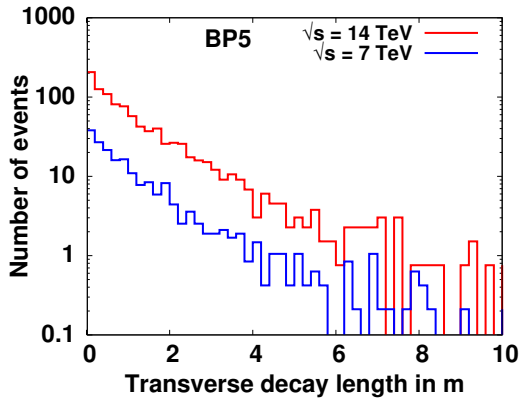}
\caption{Transverse decay length distribution of $\ntrl1$ for $\sqrt{s} = $ 7 
and 14 TeV with BP-5 for a typical detector size $\sim$ $10~m$ with $\mathcal{L}=$
$5 ~\rm{fb}^{-1}$. Minimum bin size is $20~cm$. 
The signal is given by Eq.(\ref{signal-choice}).}
\label{Decay-Length-2}
\end{figure}

For the $7$ TeV LHC run things are definitely less promising and would require a higher luminosity for claiming 
discovery. Since the average decay length is small for the two benchmark points studied for 
scenario-II (see Table \ref{decay-length2}), the fraction of events appearing outside the detector is 
$\sim 1-2\%$ of the total number of events, which is smaller compared to scenario-I. 
All of these features are also evident from Fig. \ref{Decay-Length-2}, 
where the transverse decay length distribution is shown for BP-5. So once again combining the entire range of LSP 
decay length, we can have novel final states observed not only in the muon chambers of the LHC detectors but also in 
the inner tracker and calorimeters. Therefore, with the observation of these different signatures, new 
discoveries are envisaged even with the inclusion of one-loop corrections in the scalar sector.

\begin{figure}[ht]
\centering
\includegraphics[width=5.40cm,angle=-90]{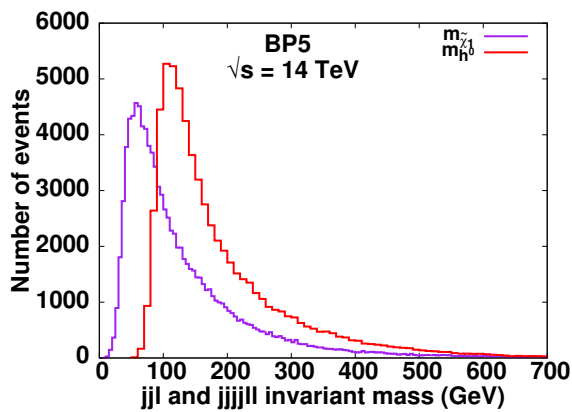}
\caption{Invariant mass distribution for $\ntrl1$ $(jj\ell )$,
as well as the Higgs boson $(jjjj\ell\ell )$. Plot is given for
$14$ TeV with $5~\rm{fb}^{-1}$
of integrated luminosity.
The Number of events for reconstructing
$m_{\ntrl1}$ for $\sqrt{s} = 14$ TeV
are scaled by a multiplicative factor $7$.}
\label{invariant-masses-2}
\end{figure}

Results of invariant mass reconstruction
for $\ntrl1$ and $h^0$ for BP-5 are shown in Fig.\ref{invariant-masses-2} 
but with $\sqrt{s}=14$ TeV of LHC run only. As already stated in the context
of Fig.\ref{invariant-masses}, we choose $jj\ell$ invariant mass 
$M(jj\ell)$ for $m_{\ntrl1}$ reconstruction.
This plot is similar to that of Fig.\ref{invariant-masses} with similar
explanations.

\section{Conclusion}\label{conclusion}
In conclusion, we have studied an unusual but spectacular signal of 
Higgs boson in supersymmetry. This signal can give rise to non-standard activities 
in the muon chamber with muon(s) and four hadronic jets. There are, however,
number of events which can leave their imprints not only at the 
muon chamber but also in the inner tracker and calorimeters concurrently. 
Integrating these two signatures can lead to discovery of an unusual signal 
of Higgs boson at the 14 TeV run of the LHC. 
Though with higher luminosity, discovery at $\sqrt{s}=$7 TeV is also possible. Indubitably, 
developement of new triggers and event reconstruction tools is essential. This signal is 
generic to a class of models where gauge-singlet neutrinos and $\rpv$ take part simultaneously 
in generating neutrino masses and mixing. Another interesting feature of this study is that 
the number of muonic events in the final state is larger than the number of electron events 
and the ratio of these two numbers can be predicted from the study of the neutrino mixing angles.
It is also important to note that such a generic conclusion remains valid with and without
considering the effect of one-loop corrections in the scalar sector. 

We thank Satyaki Bhattacharya, Manas Maity and Bruce Mellado for very helpful discussions. 
PB wants to thank  KIAS for its ``quest'' computational facility and overseas travel grant 
during the project. PG acknowledges the financial support of the CSIR, Government of India.


\end{document}